\documentclass[aps,prl,twocolumn,showpacs,superscriptaddress,groupedaddress]{revtex4}  
\usepackage{graphicx}  
\usepackage{dcolumn}   
\usepackage{bm}        
\usepackage{amssymb}   
\usepackage{amsmath}

\begin{document}


\title{ Single-Shot Readout of a Nuclear Spin Weakly Coupled to a Nitrogen-Vacancy Center}
\author {Gang-Qin Liu}
\affiliation{Beijing National Laboratory for Condensed Matter Physics, Institute of Physics, Chinese Academy of Sciences, Beijing 100190, China}
\affiliation{Department of Physics, The Chinese University of Hong Kong, Shatin, New Territories, Hong Kong, China}
\author {Jian Xing}
\affiliation{Beijing National Laboratory for Condensed Matter Physics, Institute of Physics, Chinese Academy of Sciences, Beijing 100190, China}
\author{Wen-Long Ma}
\affiliation{Department of Physics, The Chinese University of Hong Kong, Shatin, New Territories, Hong Kong, China}
\author {Chang-Hao Li}
\affiliation{Beijing National Laboratory for Condensed Matter Physics, Institute of Physics, Chinese Academy of Sciences, Beijing 100190, China}
\author{Ping Wang}
\author{Hoi Chun Po}
\author{Ren-Bao Liu}
\affiliation{Department of Physics, The Chinese University of Hong Kong, Shatin, New Territories, Hong Kong, China}
\author{Xin-Yu Pan}
\email{xypan@aphy.iphy.ac.cn}
\affiliation{Beijing National Laboratory for Condensed Matter Physics, Institute of Physics, Chinese Academy of Sciences, Beijing 100190, China}
\affiliation{Collaborative Innovation Center of Quantum Matter, Beijing 100871, China}

\date{\today}

\begin{abstract}
Single-shot readout of qubits is required for scalable quantum computing. Nuclear spins are superb quantum memories due to their long coherence times but are difficult to be read out in single shot due to their weak interaction with probes. Here we demonstrate single-shot readout of a weakly coupled $^{13}$C nuclear spin, which is unresolvable in traditional protocols. We use dynamical decoupling pulse sequences to selectively enhance the entanglement between the nuclear spin and a nitrogen-vacancy center electron spin, tuning the weak measurement of the nuclear spin to a strong, projective one. A nuclear spin coupled to the NV center with strength 330 kHz is read out in 200 ms with fidelity 95.5\%. This work provides a general protocol for single-shot readout of weakly coupled qubits and therefore largely extends the range of physical systems for scalable quantum computing.
\end{abstract}

\pacs{}
\maketitle


Nuclear spins in solids have been proposed as a promising candidate for quantum computing
\cite{kane1998silicon, Neumann2010Science_single_shot_readout,Fedor2004PRLRabi}. Quantum memory with ultra-long coherence times \cite{Lukin2012ScienceOneSecond, QuantumMemory31P_Nature2008, SSR2013Nature_Nuclear_Si}, multipartite entanglement \cite{Wrachtrup2014NatureErrorCorrection, Hanson14Error_Correction, neumann2008multipartite} and real-time feedback control \cite{Hanson2014feedback} have been demonstrated in nuclear spin systems. However, due to their small magnetic moments, it remains challenging to address and read out individual nuclear spins with high fidelity. The initialization and readout of nuclear spins around nitrogen-vacancy (NV) center in diamond, for example, require a mapping gate and an optical pulse on the NV center electron spin be repeated for more than $10^5$ times to get a sufficient signal to noise ratio \cite{Fedor2004PRLRabi, jiang2009repetitive}.

Even more challenging is single-shot readout of individual nuclear spins, which is a prerequisite of scalable quantum computing \cite{DiVincenzoQC,liuRB2010QC}. The key to single-shot readout is to acquire enough information of the quantum state before a random quantum jump event occurs.
 In previous protocols, a nearby strongly coupled electron spin is employed as an ancillary qubit.
 The nuclear spin in a state $\alpha|\downarrow\rangle + \beta|\uparrow\rangle$ can be made maximally entangled with the electron spin via a quantum control gate (C$_n$NOT$_e$), resulting in an entangled state, e.g., $\alpha|0\rangle|\downarrow\rangle + \beta|1\rangle|\uparrow\rangle$ (where
 $\{|0\rangle, |1\rangle \}$ are states of electron spin). Then the readout of the electron spin realizes a projective measurement of the nuclear spin (Pauli type-II quantum measurement).
  Since the electron spin can be rapidly initialized and read out, the process can be repeated many times before the nuclear spin undergoes a quantum jump. Therefore single-shot readout of the nuclear spin is achieved
  \cite{Neumann2010Science_single_shot_readout, SSR2011PRLFedor, Hanson11NatureElectronSSR, SSRNV2013PRL}. However, such schemes work only for nuclear spins strongly coupled to the electron spins and therefore their applications are limited to special NV centers that have $^{13}$C spins located within the first few shells. For weakly coupled nuclear spins, the hyperfine coupling is too weak to split the electron spin resonance, so it is unfeasible to apply a selective control pulse on the electron spin conditioned on the nuclear spin state, which is needed to implement an entanglement quantum gate.

\begin{figure}
\includegraphics[width=\linewidth]{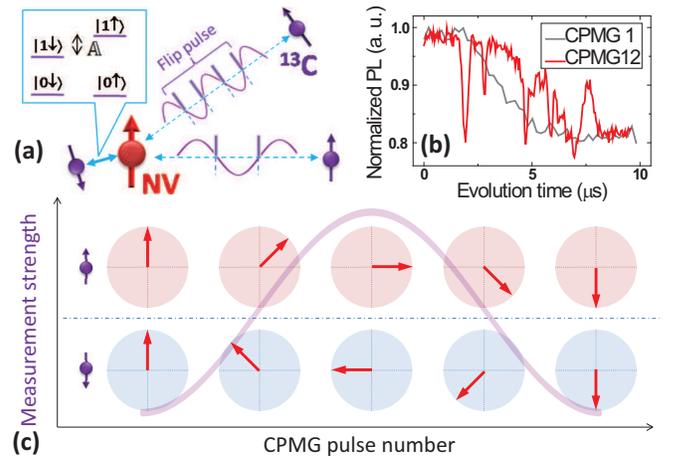}
\caption{\label{fig1} (Color online) \textbf{(color online.) System and experimental scheme for dynamical decoupling enabled quantum measurement.}
 (a) An NV electron spin (red) and its $^{13}$C nuclear spin bath (purple). The spin state of a strongly coupled $^{13}$C nuclear spin can be mapped to the electron spin through selective MW pulses. For weakly coupled nuclear spins, DD is employed to address the target one while decoupling the other nuclear spins. (b) Coherence of the center electron spin under CPMG-12 (red line) and CPMG-1 (gray line), as a function of the pulse interval $\tau$. (c) The accumulated phase of the center electron spin (and thus the measurement strength of the weakly coupled target nuclear spin) is controlled by the pulse number of applied CPMG sequence. 
 }
\end{figure}

A clue to the solution is the scheme of dynamical decoupling (DD) enabled quantum sensing of weakly coupled nuclear spins
via optically detected magnetic resonance (ODMR) of NV centers in diamond
  \cite{zhao2011Diamer, zhao2012WeakC13sensing, Lukin2012WeakC13sensing, Hanson2012WeakC13sensing, Du2014Dimer}.
  As illustrated in Fig. 1, under periodic DD control (a series of flip-flop control), the center spin can selectively accumulate phase from one weakly coupled nuclear spin (the target qubit), and the noises from other nuclear spins are suppressed.
 Furthermore, the accumulated phase from the target nuclear spin can be controlled by varying the pulse number of the DD sequence.
 As a result, the bifurcated pathways of the electron spin, which is determined by the state of the target nuclear spin, divert to a large distance and maximum entanglement between the two qubits can be established. Since the maximum entanglement is the key for projective measurement via an ancillary quantum system, we are motivated to apply the DD control for tuning the strength of quantum measurement in general and for realizing the single-shot measurement of a weakly coupled nuclear spin in particular.

The Hamiltonian of NV electron spin and a weakly coupled $^{13}$C nuclear spin $\mathbf{\hat{I}}$ under an external magnetic field \textbf{B} is \cite{ZhaoNan12PRB_decoherence,GQ13DDGate}
\begin{eqnarray}\label{}
 H = \Delta S_z^{2} + {S_z} \otimes \left( {\mathbf{A} \cdot { \hat{{\textbf{I}}}}} \right) + {\gamma _n}\mathbf{B} \cdot { \hat{{\textbf{I}}}} \nonumber,
 \label{H}
\end{eqnarray}
where the NV spin-1 $S_z$  has eigenstates  $\{|0\rangle, |\pm 1\rangle\}$, $\gamma_e=1.76\times10^{11}\mathrm{T}^{-1}\mathrm{s}^{-1}$ and  $\gamma_n=6.73\times10^{7}\mathrm{T}^{-1}\mathrm{s}^{-1}$ are the gyromagnetic ratios for the electron and nuclear spin, respectively, and $\mathbf{A}$ is the hyperfine interaction tensor. This Hamiltonian can be recast into the subspace ${|0\rangle, |-1\rangle} $ as
\begin{eqnarray}\label{}
 H = \frac{{{\sigma _z}}}{2} \otimes \mathbf{\beta}  + {H_0} \nonumber,
 \label{}
\end{eqnarray}
where $\sigma_z=|-1\rangle\langle-1|-|0\rangle\langle0|$ is the pauli operator, $\mathbf{\beta}={\mathbf{A} \cdot { \hat{{\textbf{I}}}}}$ is the noise operator, $H_0=\frac{\mathbf{A}}{2}\cdot \hat{{\textbf{I}}}+{\gamma _n}\mathbf{B} \cdot { \hat{{\textbf{I}}}}=\omega \mathbf{n}_\parallel\cdot \hat{{\textbf{I}}} $ is the effective Hamiltonian for the $^{13}$C nuclear spin.
Note that we have dropped the zero field splitting and Zeeman terms of electron spin since they have no effects in this pure-dephasing model.
 For a weakly coupled nuclear spin, consider the Carr-Purcell-Meiboom-Gill (CPMG) control with $\pi$-flips at time $t_p=(2p-1)\tau$  (where $2\tau$ is the interval between pulses and p=1,2,...N), using the Magnus expansion \cite{Ma2015sensing,albrecht2015filter}, we obtain the nuclear spin propagator conditioned on the NV electron spin state as
\begin{align}\label{}
    U_{\left(  \pm  \right)}^N\left( {t } \right) &= \exp \left( { - i\omega I_\parallel t} \right)\exp \left( { \mp \frac{{i{A_ \bot }}}{{2\omega }}F\left( {\omega ,\;t } \right){I_\perp}} \right) \nonumber,
\end{align}
where $F\left( {\omega ,\;t } \right)=|\sum^N_{p=0}\left(-1\right)^p \left( e^{-i\omega t_{p+1}}-e^{-i\omega t_p} \right)|$ is the DD filter function,
$I_{\parallel/\perp}=\mathbf{n}_{\parallel/\perp}\cdot \hat{\mathbf{I}}$,
$A_{\perp}=|\mathbf{A}-\omega \mathbf{n}_\parallel|$,
$\mathbf{n}_\perp=\left(\mathbf{A}-\omega \mathbf{n}_\parallel \right)/A_\perp$,
and the subscript +(-) denotes the propagator starts from the $|0\rangle$ ($|1\rangle$)state of the electron spin.

In general, the DD sequence steers the quantum evolution of a target nuclear spin, and the initial state and final states of the target nuclear spin under DD sequence are not the same \cite{Hanson14Error_Correction, GQ13DDGate}. Nonetheless,
we find that the eigenstates of $I_\perp$ remain in the same state after the CPMG control with an even pulse number N (see Supplementary Section 1). We denote them as $\{ |\uparrow\rangle, |\downarrow\rangle \}$, which satisfy
\begin{align}\label{}
  &U_{\left(  \pm  \right)}^N\left( {2N\tau } \right)\left|  \uparrow  \right\rangle  = {\left( { - 1} \right)^{\frac{N}{2}}}{e^{ \mp iN\phi }}\left|  \uparrow  \right\rangle \nonumber, \\
  &U_{\left(  \pm  \right)}^N\left( {2N\tau } \right)\left|  \downarrow  \right\rangle  = {\left( { - 1} \right)^{\frac{N}{2}}}{e^{ \pm iN\phi }}\left|  \downarrow  \right\rangle \nonumber,
\end{align}
where $\phi=A_\parallel/(2\omega)$ is the accumulated phase by the nuclear spin during the pulse delay $2 \tau$. Since those nuclear spin states remain the same before and after CPMG sequence, we can repetitively map and readout their quantum states with the help of the ancillary electron spin, and realize the single-shot readout of the target nuclear spin.

Besides the effect of locking the nuclear spin to the  $\{ |\uparrow\rangle, |\downarrow\rangle \}$ states, the CPMG sequence can tune the measurement strength of the target nuclear spin. The center electron spin, which is first prepared to the superposition state of $\Psi_0= \frac{1}{\sqrt{2}} \left( |0\rangle +|1\rangle \right)$, accumulates a phase determined by the state of the target nuclear spin and the pulse number of CPMG sequence, as illustrated in Fig. 1(c). After $N$ pulse CPMG sequence, the final state of electron spin is $\Psi_\uparrow= \frac{1}{\sqrt{2}} \left( |0\rangle +e^{i2N\Phi}|1\rangle \right)$ for the nuclear spin $|\uparrow\rangle$ state or $\Psi_\downarrow= \frac{1}{\sqrt{2}} \left( |0\rangle +e^{-i2N\Phi}|1\rangle \right)$ for the nuclear spin $|\uparrow\rangle$  state. In particular, if the pulse number N is such that $2 N\Phi= \pi /2 $,  the final state of the two-qubit system is $|0\downarrow\rangle$ or $|1\uparrow\rangle$ after the application of a Hadamard gate to the electron spin. Thus, a maximum entanglement between the center electron spin and the target nuclear spin can be established. A subsequent projective measurement on the electron spin will simultaneously collapse the quantum state of the target nuclear spin, which realizes a projective measurement of the target nuclear spin. As a comparison, when the pulse number is small, the nuclear spin is only weakly entangled with the ancillary electron spin, so a projective measurement of the electron spin will only cause partial collapse of the nuclear spin and hence weak measurement on the nuclear spin \cite{weakmeasurement,Hanson2014feedback}.

\begin{figure}
\includegraphics[width=\linewidth]{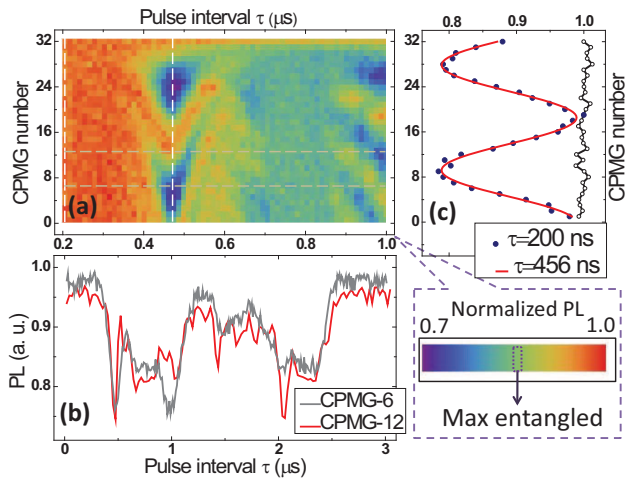}
\caption{\label{fig2} \textbf{(color online.) Controlling the strength of measurement on a weakly coupled nuclear spin by dynamical decoupling.}
(a) 2D CPMG signal as a function of the pulse number and the interval duration, under an external magnetic field of 305 Gauss. Individual nuclear spins are selected by tuning the intervals between two CPMG pulses. The number of total applied pulses determines the measurement strength of the selected nuclear spin. (b) Typical CPMG signal as a function of the interval duration. (c) Typical CPMG signal as a function of the pulse number. When no nuclear spin is resonant ($\tau$=200 ns, black line), the coherence of the center spin is well protected.
When a nuclear spin is resonantly selected ($\tau$=456 ns, red line with blue point), the coherence of the electron spin is modulated by the entanglement with the nuclear spin.
}
\end{figure}

We experimentally demonstrate our protocol on an NV center in a high-purity type-IIa diamond.
 As seen in the ODMR spectrum of the NV center (Supplementary Fig. S1), there is no apparent splitting due to hyperfine interaction with strongly coupled $^{13}$C nuclear spin. As a confirmation of the absence of strongly coupled $^{13}$C spins, the coherence in Hahn echo (CPMG-1) presents no oscillation features. Under many pulse DD control (CPMG-12), however, the measured coherence presents extended plateau and a number of dips, as shown in Fig. 1(b). So this NV center does have a few $^{13}$C nuclear spins located nearby but with weak hyperfine interaction. By fitting the CPMG signal from different initial states of the center electron spin $ \frac{1}{\sqrt{2}} \left( |0\rangle +|\pm1\rangle \right)$, we derived that the hyperfine interaction of the nearest $^{13}$C nuclear spin projected along the quantization axis of NV electron spin is about 330 kHz. Other nuclear spins have coupling strength less than this (see Supplementary Section 3).
 For all those nuclear spins ($|A| < \frac{1}{T_2^*} \approx 2 ~\mathrm{MHz}$), if taken as qubits, cannot be read out in a single shot in traditional protocols.  

The gradually enhanced entanglement between the NV center spin and a certain $^{13}$C nuclear spin is evidenced by the increasing depth of the coherence dip under more and more control pulses. As shown in Fig. 2, under an external magnetic field of 305 Gauss,
when the pulse interval matches the half precession period of an individual nuclear spin (e.g., $\tau = 456$ ns), the accumulated phase in each interval has the same direction, and the coherence dip has increasing depth with increasing the number of intervals. Actually, when the number of pulses $N$ is further increased, the over-shoot evolution of the nuclear spin can cause the disentanglement and hence the recovery of the central spin coherence. This coherence recovery effect unambiguously demonstrates the quantum nature of the noise source, i.e., the $^{13}$C nuclear spin \cite{zhao2012sensing}. Maximum entanglement between the electron and nuclear spins is reached when the pulse interval $\tau$ and the number of pulses $N$ are such that the central spin coherence exactly vanishes.
For a certain target nuclear spin (a fixed $\tau$), if it is in the fully mixed state with density matrix $\rho= 1/2$ (not polarized), the NV electron spin coherence shows oscillation behavior when the CPMG pulse number is increased \cite{Ma2015sensing}, i.e.,
\begin{align}\label{}
   {\mathcal{L}_{{\rm{dip}}}}\left( N \right) \approx \cos \left( {\frac{{{A_ \bot }N}}{\omega }} \right)\nonumber.
\end{align}
The oscillation feature of the center spin coherence presented in Fig. 2(c) is due to entanglement and disentanglement with the closest $^{13}$C nuclear spin, which has $A_\perp$= 200 kHz and $\omega = 517$ kHz under the magnetic field (305 Gauss). The measured period of coherence oscillate is $N$=16, which agree well with the theory prediction of $N= 2 \pi \omega / A_\perp$=16.3 (see Supplementary Section 3 for detail).

We utilize the oscillation of the entanglement between the electron and the nuclear spin to tune the strength of quantum measurement of the nuclear spin and realize the single-shot readout. As shown in Fig. 3(a), by tuning the pulse number of CPMG sequence, a maximum entanglement between the electron spin and the weakly coupled nuclear spin is prepared. A subsequent projection measurement of the electron spin, which is realized by a short optical pulse, causes the quantum state of the target nuclear spin to collapse. After the first cycle of measurement on the electron spin, the nuclear spin randomly collapses to $|\downarrow\rangle$ or $|\uparrow\rangle$, and in subsequent cycles of measurement, it will stay on the same state until a quantum jump occurs. Since the relaxation time (T$_1$) of the $^{13}$C nuclear spin is much longer than the CPMG sequence and the optical readout duration ( $\approx\mu$s), we can repeat many cycles of measurement on the electron spin to accumulate sufficient statistical confidence before the nuclear spin undergoes a random quantum jump. Note the readout result is determined by the first random collapse of the nuclear spin state, so it is a single-shot readout of the target nuclear spin.

\begin{figure}
\includegraphics[width=\linewidth]{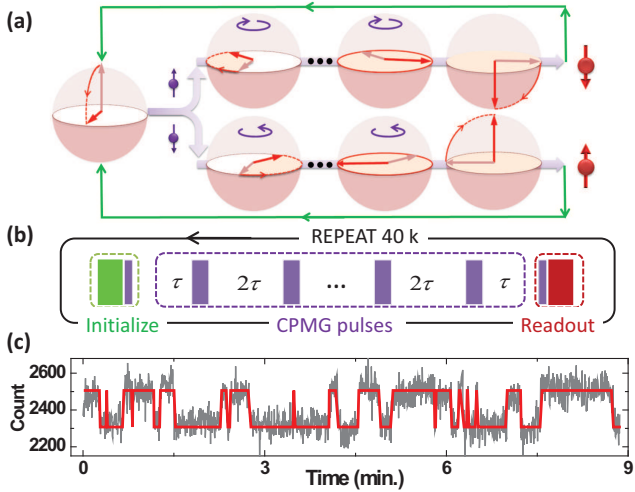}
\caption{\label{fig3} \textbf{(color online.) Single-shot readout of a weakly coupled nuclear spin.}
 (a) Bifurcated pathways of the electron spin under DD-enabled single-shot readout of a weakly coupled nuclear spin. (b) Pulse sequence of DD-enabled single-shot readout. (c) Photon count trace for the pulse sequence depicted in (b) with CPMG-12, $\tau$=248 ns, under an external magnetic field of 691 Gauss. Each data point is a sum of 40,000 cycles of measurement of the center electron spin. The quantum jump behavior indicates that the target nuclear spin (with projected hyperfine interaction of $A_\perp$=255 kHz under this magnetic field) is efficiently initialized and read out by the single shot readout sequence.
}
\end{figure}

Under an external magnetic field of 691 Gauss, we use a CPMG-12 sequence with $\tau =$ 248 ns and a following optical pulse (300 ns) to read out the closest nuclear spin (corresponds to the first coherence dip in Fig. 1(b)). A short waiting interval is added after the readout and re-initialization of electron spin to make the measurement duration matches the precession period of target nuclear spin ( $\approx 1~\mu$s under this magnetic field), see Fig. 3(b) for the pulse sequence. The photon counts in 40,000 cycles (in 189 ms) are summed together to a single data point, which presents the state of the nuclear spin under measurement.

With the single-shot measurement, we are able to directly observe the quantum jumps of the target nuclear spin. Typical photon count trace is shown in Fig. 3(c). Two distinct values are clearly seen. We associate the high (low) count rate to the $|\uparrow\rangle$ ($|\downarrow\rangle$) state of the target nuclear spin. Under the magnetic field of B= 691 Gauss along the quantization axis of this NV center, a relaxation time ($T_{1n}$) of about 15 second for both $|\uparrow\rangle$ and $|\downarrow\rangle$ states is measured. The initialization and readout fidelity of our single-shot readout protocol is estimated as following \cite{SSRNV2013PRL}. First, two photon count thresholds are introduced to distinguish the nuclear spin $|\uparrow\rangle$ and $|\downarrow\rangle$ states, being above 2520 and below 2300, respectively. This way, we prepare the nuclear spin state with a fidelity $>$99\% (for both states) in a single-shot measurement. After that, a subsequent single-shot readout is performed, from which the photon count distributions dependent on the nuclear spin state are extracted. As shown in Fig. 4(a), the photon count distributions for the nuclear spin state being initialized to the $|\uparrow\rangle$ and $|\downarrow\rangle$ states are well distinguished. Then we calculate the readout fidelity by comparing the possibility of successful readout and total readout, in an integration interval determined by the photon count threshold \cite{Neumann2010Science_single_shot_readout,Lukin2012ScienceOneSecond, SSRNV2013PRL}. Fig. 4(b) presents the readout fidelity as a function of the threshold. For a threshold of 2400, which is the maximum overlap between the photon counting distributions for the two nuclear spin states, the readout fidelity is 95.5\% (for both $|\uparrow\rangle$ and $|\downarrow\rangle$  states). The fidelity can be further improved by increasing the photon collection efficiency and the fidelity of flip pulses.

The DD-enabled single-shot readout of a weakly coupled nuclear spin largely extends the range of physical systems for scalable quantum computing. This is especially useful for individually addressing each nuclear qubit in an NV centers weakly coupled to a number of nuclear spin qubits. The nuclear spins can be read out in succession, which can be employed to implement measurement-based entanglement between two nuclear spins, even if there is no direct interaction between them \cite{EntangleByMeas_NV}. Meanwhile, as demonstrated in our previous work \cite{GQ13DDGate}, the quantum evolution of a target nuclear spin can be steered by an engineered DD sequence, thus all key elements of quantum computing, including initialization, manipulation, and readout, can be realized by DD sequences, while the coherence of the center electron spin is well protected.

\begin{figure}
\includegraphics[width=\linewidth]{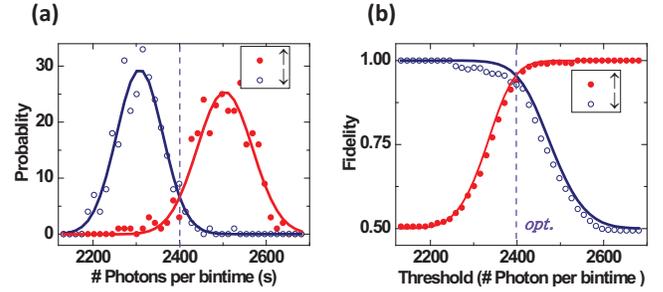}
\caption{\label{fig4} \textbf{(color online.) Fidelity of the single shot readout of a weakly coupled nuclear spin.}
  (a)Photon count distributions after the target nuclear spin state selected by the photon count thresholds of $<$2300 for $|\downarrow\rangle$ and $>$2520 for $|\uparrow\rangle$. The distributions of the nuclear spin state after being initialized to the $|\downarrow\rangle$ and $|\uparrow\rangle$ states are well distinguished. (b) Single-shot readout fidelity as a function of the readout threshold. With a threshold of 2400, the readout fidelity is 95.5\% for both $|\downarrow\rangle$ and $|\uparrow\rangle$.
}
\end{figure}

In conclusion, we propose and demonstrate a scheme of single-shot readout of a weakly coupled nuclear spin by applying dynamical decoupling on an ancillary electron spin. The protocol can extract the state information of a target nuclear spin while keeping the other nuclear spins untouched. For the selected target nuclear spin, the measurement strength is tunable by adjusting the CPMG number, thus both strong and weak measurement can be achieved for the same system. Weak measurement of a quantum system is of particular interest since the target system can be steered through the back action of sequential weak measurements and real-time feedback \cite{Hanson2014feedback, quantum_trajectories, weber2014mapping}, together with the demonstrated strong measurement, it is possible to demonstrate measurement-only quantum computing.


\emph{Acknowledgements} This work was supported by National Basic Research Program of China (973 Program project No. 2014CB921402), the Strategic Priority Research Program of the Chinese Academy of Sciences (under Grant No. XDB07010300), Hong Kong Research Grants Council - Collaborative Research Fund Project CUHK4/CRF/12G, and The Chinese University of Hong Kong Vice Chancellor's One-off Discretionary Fund.


\end{document}